\documentclass[10pt,twoside]{article}
\usepackage{graphicx}
\usepackage{amsmath}
\usepackage{Latex-document}
\newcommand\from{:}

\newcommand\card[1]{\left| #1 \right|}
\newcommand\set[1]{\left\{#1\right\}}
\newcommand\poly{\mbox{poly}}

\newcommand\p{\mbox{P}}
\newcommand\pcp{\mbox{PCP}}

\newcommand\np{\mbox{NP}}

\newcommand\cC{{\cal C}}

\newcommand\mindist{\mbox{min-dist}}

\title{\bf How NP Got a New Definition: A Survey \vskip -2mm
of Probabilistically Checkable Proofs\thanks{Supported by David
and Lucile Packard Fellowship, NSF Grant CCR-0098180, and an NSF
ITR Grant.}\vskip 6mm}

\author{Sanjeev Arora\vspace*{-0.5cm}\thanks{Department of Computer Science,
Princeton University, Princeton, NJ 08544, USA. Email: arora@cs.princeton.edu,  Web page:
www.cs.princeton.edu/\~{}arora/}}

\date{\vspace{-8mm}}

\begin{document}

\maketitle

\thispagestyle{first} \setcounter{page}{637}

\begin{abstract}

\vskip 3mm

We survey a  collective achievement of a group of researchers:
the PCP Theorems. They give
new definitions of the class \np, and imply that computing
approximate solutions to many \np-hard problems is itself \np-hard.
Techniques developed to prove them have had many other consequences.

\vskip 4.5mm

\noindent {\bf 2000 Mathematics Subject Classification:} 68Q10,
68Q15, 68Q17, 68Q25.

\noindent {\bf Keywords and Phrases:} Complexity theory, NP,
Probabilistically checkable proofs, Approximation algorithms.
\end{abstract}

\vskip 12mm

\section{PCP theorems: an informal introduction} \label{sec:intro}\setzero
\vskip-5mm \hspace{5mm}

Suppose a mathematician circulates a proof of an important result, say
Riemann Hypothesis, fitting several thousand pages. To verify it would
take you and your doubting colleagues
several years. Can you do it faster?
Yes, according to the PCP Theorems. He can
rewrite his proof so you can verify it by probabilistically
selecting (i.e., using a source of random bits)  a constant number of bits
---as low as $3$ bits---to examine in it. Furthermore, this verification
has the following properties: (a) A correct proof will never fail
to convince you (that is, no choice of the random bits will make
you reject a correct proof) and (b) An incorrect proof will
convince you with only negligible probability ($2^{-100}$ if you
examine $300$ bits). In fact, a stronger assertion is true: if the
Riemann hypothesis is false, then you are guaranteed to reject
{\em any string of letters placed before you} with high
probability after examining a constant number of bits. (c) This
proof rewriting is completely mechanical---a computer could do
it---and does not greatly increase its size. ({\em Caveat}: Before
journal editors rush to adopt this new proof verification, we
should mention that it currently requires  proofs written in a
formal axiomatic system
---such as Zermelo Fraenkel set theory---since computers do not
 understand English.)

This result has a strong ring of implausibility.
A mathematical proof is invalid if it has even a single error somewhere.
How can this error spread itself all over the rewritten proof,
so as to be apparent after
we have probabilistically examined a few bits in the proof?
(Note that the simple idea of just making multiple copies of
the erroneous line everywhere
does not work: the unknown mathematician could hand you a proof
 in which this does not happen, yet that does not make the proof correct.)
The methods used to achieve this level of redundancy
are reminiscent of the theory of
error-correcting codes, though they are novel and interesting in their own
right, and their full implications are still being felt (see Section~\ref{sec:proof}).

\subsection{New definition of \np}
\vskip -5mm \hspace{5mm}

The PCP Theorems  provide interesting new definitions for the
complexity class \np. (Clarification: the singular form ``PCP
Theorem'' will refer to a single result $\np =\pcp(\log n, 1)$
proved in \cite{arsa,ALMSS}, and the plural form ``PCP Theorems''
refers to a  large body of results  of a similar ilk, some
predating the PCP Theorem.) Classically, \np\ is defined as the
set of decision problems for which a ``Yes'' answer has a short
certificate verifiable in polynomial time (i.e., if the instance
size is $n$, then the  certificate size and the verification time
is $n^c$ for some fixed constant $c$). The following are two
examples:

\vskip 1mm

\noindent 3-SAT $=$ satisfiable boolean formulae of the form AND of
clauses of size at most $3$, e.g.,
$(x_1 \vee \neg x_2 \vee x_3) \wedge (\neg x_1 \vee x_2 \vee x_3) \wedge (x_4)$.
(The certificate for  satisfiability is simply
an assignment to the variables that makes the formula true.)

\vskip 1mm

\noindent $\mbox{MATH-THEOREM}_{ZFC}$ $=$ set of strings of the form
$(T, 1^n)$ where
$T$ is a mathematical statement that is a theorem in Zermelo Fraenkel set theory
that has a proof $n$ bits long. (The ``certificate'' for theoremhood
is just the proof.)

The famous conjecture $\p \neq \np$ ---now  one of seven Millenium Prize
problems in math~\cite{clayprobs}---says that not every \np\ problem is
solvable in polynomial time. In other words,
though the certificate is easy
to check, it is not always easy to find.

The PCP Theorem gives  a new definition of \np: it is the set of
decision problems for which a ``Yes'' answer has a polynomial-size
certificate which can be probabilistically checked using $O(\log n)$ random
bits and by examing $O(1)$ (i.e., constant) number of bits in it.

Our earlier claim about  proof verification follows from the PCP Theorem,
since $\mbox{MATH-THEOREM}_{ZFC}$ is in  $\np$,
and hence there is a way to certify a YES answer (namely,
theoremhood) that
satisfies properties (a) and (b).
(Property (c) follows from the
constructive nature of the proof of the PCP Theorem in \cite{arsa,ALMSS}.)

Motivated by the PCP Theorems, researchers have proved new
analogous definitions of other complexity classes
such as PSPACE~\cite{cfls2} and PH~\cite{klrss}.

\subsection{Optimization, approximation, and PCP theorems}
\vskip -5mm \hspace{5mm}

 The $\p$ versus $\np$ question is
important because of  {\em $\np$-completeness} (also, \np-{\em
hardness}). Optimization problems in a variety of disciplines are
\np-hard~\cite{GarJon}, and so if $\p \neq \np$ they cannot be
solved in polynomial time. The following is one such optimization
problem.

%\medskip

\noindent MAX-3SAT: Given a 3-CNF boolean formula $\varphi$, find an assignment
to the variables that maximizes the number of satisfied clauses.

{\em Approximation algorithms} represent a way to deal with \np-hardness.
An algorithm {\em achieves an approximation ratio\/}
$\alpha$ for a maximization  problem if, for {\em every\/} instance,
it produces a solution of value  at least $OPT/\alpha$, where $OPT$ is
the value of the optimal solution. (For a minimization problem,
achieving a ratio $\alpha$ involves finding a solution of
cost at most $\alpha \,OPT$.) Note that the approximation
ratio is $\geq 1$ by definition. For MAX-3SAT we now know a
polynomial-time algorithm that achieves an approximation ratio $8/7$~\cite{KZ}.

Though approximation algorithms is a well-developed
research area (see~\cite{hochbook,vijaybook}),
for many problems no good approximation algorithms have been found.
The PCP Theorems suggest
a reason: for many NP-hard problems, including
MAX-CLIQUE, CHROMATIC NUMBER, MAX-3SAT, and SET-COVER,
achieving certain reasonable  approximation ratios is
no easier than computing optimal solutions. In other words,
approximation is NP-hard. For instance, achieving a ratio
$8/7 -\epsilon$ for MAX-3SAT is \np-hard~\cite{hastadmtsat}.
%Proving such results was a major impetus in the
%discovery of the PCP Theorems.

Why do the PCP Theorems lead to   such results? Details appear
in the survey~\cite{arlu} (and [Feige 2002], these proceedings), but
we hint at  the reason
using 3SAT and  MAX-3SAT as examples.
Cook and Levin~\cite{cook,levin} showed how to  reduce any \np\ problem
to 3SAT, by constructing, for any nondeterministic
machine, a  3CNF formula whose satisfying assignments
represent the  transcripts of accepting computations.
%Thus every \np\ problem
%reduces in polynomial time to 3SAT.
Thus it is difficult to satisfy all clauses.
Yet it is easy to find assignmenent
satisfying $1 -o(1)$ fraction of the clauses!
The reason is that a computation transcript is a very {\em non-robust\/}
object: changing even a bit affects its correctness.
%obtained from  Cook-Levin
%are {\em almost satisfiable}:
Thus  the Cook-Levin reduction does not prove the inapproximability of
MAX-3SAT. By providing a more
robust representation of a computation, the PCP Theorems
 overcome this difficulty.
We note that  MAX-3SAT is a central problem in
the study of inapproximability: once we have proved its inapproximability,
other inapproximability results easily follow (see \cite{arlu};
the observation in a weaker form is originally from work on
MAX-SNP~\cite{PY}).

%In a survey article, Arora and Lund~\cite{arlu} briefly describe
%how to prove most known inapproximability results. They list at least two dozen
%important problems for which inapproximability was
%proved using \pcp-based techniques. They make the empirical observation that
%these problems divide into four broad classes, based upon the
%approximation ratio that is provably hard to achieve for them.
%Class I contains all problems for which achieving
%an approximation ratio $1+\epsilon$
%is \np-hard for some fixed $\epsilon >0$.  Classes II, III, and
%IV contain problems for which the corresponding ratios
%are $\Theta(\log n)$, $2^{\log^{1-\epsilon}n}$ for every fixed
%$\epsilon >0$, and $n^{\epsilon}$ for some fixed $\epsilon >0$.
%(Of course, each class contains every  higher numbered class.)
%Inapproximability results within a class seem to use similar
%techniques, and in fact the authors identify a canonical problem in each
%class which can be used to
%prove the inapproximability results for all other problems in that
%class. The inapproximability of the canonical problems can be
%proved using MAX-3SAT.
%The authors pose an open question whether or not
%these empirically observed classes can be derived from
%some deeper theory. (Only Class I seems to have an
%explanation, using

\subsection{History and context}
\vskip -5mm \hspace{5mm}

 PCPs evolved from {\em interactive
proofs}, which were invented by Goldwasser, Micali, and
Rackoff~\cite{GMR} and Babai~\cite{babai} as a probabilistic
extension of NP and proved useful in cryptography and complexity
theory (see Goldreich's survey~\cite{G}), including some early
versions of PCPs~\cite{FRS}. In 1990, Lund, Fortnow, Karloff and
Nisan~\cite{LFKN} and Shamir~\cite{Sha} showed IP=PSPACE, thus
giving a new probabilistic definition of PSPACE in terms of
interactive proofs. They introduced a revolutionary algebraic way
of looking at boolean formulae. In restrospect, this
algebraization can also be seen as a ``robust'' representation of
computation.
%(cf.~Section~\ref{sec:history}).
The inspiration to use polynomials
came from works on {\em program checking}~\cite{BK} (see also
\cite{lipton,befe,BLR}).
Babai, Fortnow, and Lund~\cite{BFL} used similar methods to give a
new probabilistic definition of NEXPTIME, the exponential analogue of
\np. To extend this result to \np, Babai, Fortnow, Levin, and
Szegedy~\cite{BFLS} and Feige, Goldwasser, Lov{\'a}sz, Safra, and
Szegedy~\cite{FGLSS} studied variants of what we now call
probabilistically checkable proof
systems (Babai et al.~called their systems {\em holographic\/}
proofs).

Feige et al.~also proved the first inapproximability result
in the PCP area: if any polynomial-time algorithm can achieve a
constant approximation
ratio for the MAX-CLIQUE problem, then every $\np$ problem is solvable
in $n^{O(\log \log n)}$ time. This important result
drew  everybody's attention to the (as yet unnamed)
area of probabilistically checkable proofs.
A year later, Arora and Safra~\cite{arsa}
formalized and named the class PCP and used it to give a new probabilistic
definition of \np. (Babai et al. and Feige et al.'s results were
precursors of this new definition.)  They also showed that approximating
MAX-CLIQUE is \np-hard.
Soon, Arora, Lund, Motwani, Sudan, and Szegedy~\cite{ALMSS} proved
the PCP Theorem (see below) and showed that MAX-SNP-hard problems do not have a PTAS
if $\p \neq \np$. Since the second paper relied heavily on the still-unpublished
first paper, the the PCP theorem is
jointly attributed to \cite{arsa,ALMSS}. For surveys of these
developments see~\cite{babaisurvey,G,johnsonsurvey,MPS}.
%In the years since the discovery
%of the PCP Theorem,
%other variants of  PCP have been studied  and used in
% inapproximability results, and we will refer to them collectively as
%PCP Theorems.

%The major PCP Theorems are in ALMSS, Hastad, Raz.

\section{Definitions and results} \label{sec:PCP}

\vskip-5mm \hspace{5mm}

Now we define the class \pcp. We will use ``language membership'' and ``decision
problem'' interchangeably.
A $(r(n), q(n))$-{\em restricted verifier} for a language $L$,
where $r, q$ are integer-valued functions, is a probabilistic turing machine
$M$ that, given an input of size $n$, checks membership certificates for the input
in the following way. The certificate is an array of bits to which
the verifier has random-access (that is, it
can {\em query\/} individual bits of the certificate).
\begin{itemize}
\item The verifier reads the input, and
uses $O(r(n))$ random bits to compute a sequence of $O(q(n))$ addresses in
the certificate.
\item The verifier queries the bits at those addresses, and depending upon
what they were, outputs ``accept'' or ``reject''.
\item \begin{equation}
 \label{eqn:complete}
\forall x \in L~~\mbox{$\exists$ certificate $\Pi$ s.t.
$\Pr[M^{\Pi}\mbox{accepts}] =1$},
\end{equation}
\begin{equation}
\label{eqn:sound}
\forall x \not \in L~~\mbox{$\forall$ certificate $\Pi$,
$\Pr[M^{\Pi}\mbox{accepts}] \leq 1/2$}
\end{equation}
(In both cases the probability is over the choice of the verifier's random
string.)
\end{itemize}
$\pcp(r(n), q(n))$ is the complexity class consisting of every language with an \linebreak $(r(n),
q(n))$-restricted verifier. Since \np\ is the class of languages whose membership certificates can be checked by a
deterministic polynomial-time verifier, $\np =\cup_{c\geq 0}\pcp(0, n^c)$. The PCP Theorem gives an alternative
definition:
%\begin{equation}
$\np =\pcp(\log n, 1)$.
%\end{equation}
Other \pcp-like classes have been defined by using variants of
the definition above, and shown to equal \np\ (when the  parameters
are  appropriately chosen). We mention some variants
and the best results known for them; these are the ``PCP Theorems'' alluded to earlier.

\begin{enumerate}
\item The probability  $1$ in condition~(\ref{eqn:complete}) may be
allowed to be  $c <1$. Such a verifier is said to have
{\em imperfect completeness} $c$.
\item The probability  $1/2$ in condition~(\ref{eqn:sound}) may be allowed to
be $s <c$. Such a verifier is said to have {\em soundness} $s$.
Using standard results on
random walks on expanders, it can be shown from the PCP theorem
that every \np\ language has verifiers with perfect completeness
that use $O(k)$ query bits for soundness $2^{-k}$ (here $k\leq O(\log n)$).
\item The number of query bits, which was $O(q(n))$ above, may be specified
more precisely together with the leading constant. The constant is important
for many inapproximability results.
Building upon past results on PCPs and using fourier analysis,
H{\aa}stad~\cite{hastadmtsat} recently proved
that for each $\epsilon >0$, every \np\
language has a verifier with completeness $1-\epsilon$, soundness
$1/2$ and only $3$ query bits. He uses this to show the
inapproximability of MAX-3SAT upto a factor $8/7 -\epsilon$.
\item The {\em free bit} parameter may be used instead of
query bits~\cite{FKcli,BS}.
This parameter is defined as follows. Suppose the query bit
parameter is $q$. After the
verifier has picked its random string, and picked a sequence of $q$ addresses,
there are $2^q$ possible sequences of bits that could be contained in
those addresses.
If the verifier accepts for only $t$ of those sequences, then we say that
the free bit parameter is $\log t$ (note that this number need not be
an integer). Samorodnitsky and Trevisan show how to reduce the soundness
to $2^{-k^2/4}$ using $k$ free bits~\cite{ST}.
\item {\em Amortized free bits} may be used~\cite{BS}. This parameter
is $\lim_{s \to 0} f_s/\log (1/s)$, where
$f_s$ is the number of free bits needed by the verifier to make
soundness $<s$. H{\aa}stad~\cite{hastadcli}
shows that for each $\epsilon >0$,
every \np\ language has a verifier that uses $O(\log n)$ random bits
and $\epsilon$ amortized free bits.
He uses this to show (using a reduction from~\cite{FGLSS} and
modified by \cite{FKcli,BS}) that MAX-CLIQUE is inapproximable
upto a factor $n^{1-\delta}$.
\item The certificate may contain not bits
but letters from a larger alphabet $\Sigma$. The verifier's
soundness may then depend upon $\Sigma$.
In a {\em $p$ prover 1-round interactive proof system},
the certificate consists of $p$  arrays of letters from $\Sigma$.
The verifier is only allowed to query $1$ letter from each array.
Since each letter of $\Sigma$ is represented by
$\lceil \log \card{\Sigma} \rceil$
bits, the  number of {\em bits} queried may be viewed as
$p\cdot\lceil \log \card{\Sigma} \rceil$.
Constructions of such proof systems for \np\ appeared in
\cite{BGKW,LS,FL,BGLR,FKcli,raz}. Lund and Yannakakis~\cite{LY} used these
proof systems to prove inapproximability results for
SETCOVER and many subgraph maximization problems.
The best construction of such proof systems is due to
Raz and Safra~\cite{razsaf}. They
show  that for each $k \leq \sqrt{\log n}$,
every \np\ language has  a verifier that uses
$O(\log n)$ random bits, has $\log \card{\Sigma} = O(k)$ and soundness
$2^{-k}$. The parameter $p$ is $O(1)$.
\end{enumerate}

\section{Proof of the PCP theorems} \label{sec:proof}

\vskip-5mm \hspace{5mm}

A striking feature of the PCP Theorems is that each builds upon the
previous ones.
%Consequently,
%a proof of H{\aa}stad's MAX-CLIQUE
%result from first principles would fill well over 100 pages!
However, a few ideas recur.
First, note that it suffices to design verifiers for
3SAT since 3SAT is \np-complete and a verifier for any other language can
transform the input to a 3SAT instance as a first step.
The verifier then expects a certificate for a ``yes'' answer
to be an encoding of a satisfying assignment; we define this next.

For an alphabet $\Sigma$ let $\Sigma^m$ denote the set of $m$-letter words.
The {\em distance} between two words $x, y \in \Sigma^m$,
denoted $\delta(x, y)$, is the fraction
of indices they differ on. For a set $\cC \subseteq \Sigma^m$, let
the {\em minimum distance} of $\cC$, denoted $\mindist(\cC)$, refer to
$\min_{x, y \in \cC; x \neq y}\set{\delta(x, y)}$
and let $\delta(x, \cC)$ stand for $\min_{y \in \cC}\set{\delta(x, y)}$.
If $\mindist(\cC) =\gamma$, and $\delta(x, \cC) <\gamma/2$, then triangle inequality implies
there is a unique $y \in \cC$ such that $\delta(x, y) = \delta(x, \cC)$.
We will be interested in $\cC$ such that $\mindist(\cC)\geq 0.5$; such
sets are examples of {\em error-correcting codes} from information theory,
where $\cC$ is thought of as a map from strings of $\log \card{\cC}$ bits (``messages'')
to $\cC$. When encoded this way, messages can be recovered even
if transmitted over a noisy channel
that corrupts up to $1/4$th of the letters.

The probabilistically checkable certificate is required to contain the
encoding of a satisfying assignment using some such $\cC$.
When presented with such a string, the verifier needs to check, first, that
the string is {\em close} to some codeword, and second, that the (unique) closest
codeword is the encoding of a satisfying assignment.
As one would expect, the set $\cC$ is defined using mathematically interesting objects
(polynomials, monotone functions, etc.) so the final technique may be seen as
a ``lifting'' of the satisfiability question to some mathematical domain (such as
algebra). The important new angle is ``local checkability,'' namely, the
ability to verify global properties by a few random spot-checks. (See below.)

Another important technique introduced in~\cite{arsa} and used in all
subsequent papers is {\em verifier composition}, which composes two
verifiers to give a new verifier some of whose parameters are lower
than those in either verifier. Verifier composition relies on
the notion of a {\em probabilistically checkable split-encoding},
a notion to which Arora and Safra were led by results in \cite{BFLS}.
(Later PCP Theorems use other probabilistically checkable encodings:
{\em linear function codes}~\cite{ALMSS},  and {\em long codes}~
\cite{BGS,hastadcli,hastadmtsat}.)
One final but crucial ingredient in recent PCP Theorems
is Raz's {\em parallel repetition theorem}~\cite{raz}.

\subsection{Local tests for global properties}
\vskip -5mm \hspace{5mm}

 The key idea in the PCP Theorems is to
design probabilistic local checks that verify  global properties
of a provided certificate. Designing such local tests involves
proving a statement of the following type: if a certain object
satisfies some local property ``often'' (say, in $90\%$ of the
local neighborhoods) then it satisfies a global property. Such
statements are reminiscent of theorems in more classical areas of
math, e.g., those establishing properties of  solutions to PDEs,
but the analogy is not exact because we only require the local
property to hold in most neighborhoods, and not all.

We illustrate with some examples. (A research area called
{\em Property Testing}~\cite{dron} now consists of  inventing such local tests
for different properties.)
There is a set $\cC \subseteq \Sigma^m$
of interest, with $\mindist(\cC) \geq 0.5$. Presented with $x \in \Sigma^m$, we
wish to read ``a few'' letters in it to determine whether
$\delta(x, \cC)$ is small.

\begin{enumerate}
\item {\em Linearity test}. Here $\Sigma = GF(2)$ and $m=2^n$ for some integer $n$.
Thus $\Sigma^m$ is the set of all functions from $GF(2)^n$ to $GF(2)$.
Let $\cC_1$ be the set of words that correspond to {\em linear functions}, namely,
the set of $f \from GF(2)^n \to GF(2)$ such that $\exists a_1,\ldots, a_n \in GF(2)~\mbox{s.t.} f(z_1, z_2, \ldots,, z_n)= \sum_i a_i z_i.$
The test for linearity involves picking $\overline{z}, \overline{u} \in GF(2)^n$
randomly and accepting iff $f(\overline{z}) + f(\overline{u}) = f(\overline{z}+
\overline{u})$. Let $\gamma$ be the probability that this test does not accept.
Using elementary fourier analysis one can show
$\gamma \geq \delta(f, \cC_1)/2$~\cite{bchks} (see also earlier weaker results
in \cite{BLR}).
\item {\em Low Degree Test}. Here $\Sigma =GF(p)$ for a prime $p$ and $m = p^n$
for some $n$. Thus $\Sigma^m$ is the set of all functions from $GF(p)^n$ to $GF(p)$.
Let $\cC_2$ be the set of words that correspond to
{\em polynomials of total degree $d$}, namely, the set of
$f \from GF(p)^n \to GF(p)$ such that
there is a $n$-variate polynomial $g$ of degree $d$ and
$f(z_1, z_2, \ldots,, z_n)= g(z_1, z_2, \ldots,, z_n).$
We assume $dn \ll p$ (hence degree is ``low'').
Testing for closeness to $\cC_2$ involves
picking random lines. A line has
the parametric form $\set{(a_1 + b_1t, a_2 + b_2 t, \ldots, a_n + b_n t): t \in GF(p)}$
for some $a_1, a_2, \ldots, a_n$, $b_1, b_2,\ldots, b_n \in GF(p)$. (It is
a $1$-dimensional affine subspace, hence much smaller than $GF(p)^n$.)
Note that if $f$ is described by a degree $d$ polynomial, then its restriction to
such a line is described by a univariate degree $d$ polynomial in the line parameter $t$.

\vskip 1mm

\begin{itemize}
\item Variant 1: Pick a random line, read its first $d+1$ points  to construct a degree $d$
univariate polynomial, and check if it describes $f$ at a randomly chosen point of
the line. This test appears in \cite{RS} and is similar to another test in \cite{FGLSS}.
%Let $\gamma$ be the probability that the test does not accept. Then
%$\gamma \geq \Omega(\delta(f, \cC_2)/d)$.
\item Variant 2: This test uses the fact that in the PCP setting, it is
reasonable to ask
that the provided certificate should contain additional useful information to facilitate
the test. We require, together with $f$, a separate  table containing a degree $d$
univariate polynomial for the line. We do the test above, except after
picking the random line
we read the relevant univariate polynomial from the provided table. This has the
crucial benefit
that we do not have to  read $d+1$ separate ``pieces'' of information from the
two tables.
If $\gamma$ is the probability that the test rejects, then
$\gamma \geq \min\set{0.1, \delta(f , \cC_2)/2}$  (see~\cite{ALMSS};
which uses  \cite{RS,arsa}).
\end{itemize}

\item {\em Closeness to a small set of codewords}. Above, we wanted to check
that $\delta(f, \cC) <0.1$, in which case there is a unique word from
$\cC$ in $\mbox{Ball}(f, 0.1)$. Proofs of recent PCP Theorems relax this and
only require for some $\epsilon$
that there is a small set of words $S \subseteq \cC$
such that each $s \in S$ lies in $\mbox{Ball}(f, \epsilon)$. (In information
theory, such an $S$ is
called a {\em list decoding} of $f$.) We mention two important such tests.

\medskip

\noindent{For degree $d$ polynomials}:  The test in Variant 2 works with
a stronger guarantee: if $\beta$ is the probability that the test accepts, then
there are $\poly(1/\epsilon)$ polynomials whose distance to $f$ is less than
$1-\epsilon$ provided $p > \poly(nd/\beta \epsilon)$ (see~\cite{arsu}, and also
\cite{razsaf} for an alternative test).

\medskip

\noindent {\em Long Code test}. Here $\Sigma = GF(2)$ and $m=2^n$ for some integer $n$.
Thus $\Sigma^m$ is the set of all functions from $GF(2)^n$ to $GF(2)$.
Let $\cC_3$ be the set of words that correspond to {\em coordinate functions},
namely,
$$\set{f \from GF(2)^n \to GF(2):~\exists i \in \set{1, 2,\ldots, n}~\mbox{s.t.}
f(z_1, z_2, \ldots, z_n)= z_i.}$$ (This encodes $i\in [1,n]$,
i.e., $\log n$ bits of information, using a string of length
$2^{n}$, hence the name ``Long Code''.) The following test
works~\cite{hastadmtsat}, though we do not elaborate on the exact
statement, which is technical: Pick $\overline{z},\overline{w} \in
GF(2)^n$ and $\overline{u}\in GF(2)^n$ that is a random vector
with $1$'s in $\epsilon$ fraction of the entries. Accept iff
$f(\overline{z}+\overline{w}) = f(\overline{z})+ f(\overline{w}+
\overline{u})$. (Note the similarity to the linearity test above.)
\end{enumerate}

\subsection{Further applications of PCP techniques}
\vskip -5mm \hspace{5mm}

We list some notable applications of PCP techniques. The PCP
Theorem is  useful in cryptography because many cryptographic
primitives involve basic steps that prove Yes/No assertions that
are in  \np (or even \p). The PCP Theorem allows this to be done
in a communication-efficient manner. See
\cite{kilian,micaliCS,barak} for some examples. Some stronger
forms of the PCP Theorem (specifically, a version involving
encoded inputs) have found uses in giving new definitions for
polynomial hierarchy~\cite{klrss} and PSPACE~\cite{cfls,cfls2}.
Finally, the properties of polynomials and polynomial-based
encodings discovered for use in PCP Theorems have influenced  new
decoding algorithms for error-correcting codes~\cite{GS},
constructions of pseudorandom graphs called {\em
extractors}~\cite{trevisan,STZ} and derandomization techniques in
complexity theory (e.g.~\cite{STV}).

\newcommand{\etalchar}[1]{$^{#1}$}

\newcommand{\stoc}[1]{\ifnum#1=
71{{\sl Proceedings of the Third Annual Symposium
on the Theory of Computing,\/} ACM, 1971}\else{\ifnum#1=
72{{\sl Proceedings of the Fourth Annual Symposium
on the Theory of Computing,\/} ACM, 1972}\else{\ifnum#1=
83{{\sl Proceedings of the Fifteenth Annual Symposium
on the Theory of Computing,\/} ACM, 1983}\else{\ifnum#1=
84{{\sl Proceedings of the Sixteenth Annual Symposium
on the Theory of Computing,\/} ACM, 1984}\else{\ifnum#1=
85{{\sl Proceedings of the Seventeenth Annual Symposium
on the Theory of Computing,\/} ACM, 1985}\else{\ifnum#1=
86{{\sl Proceedings of the Eighteenth Annual Symposium
on the Theory of Computing,\/} ACM, 1986}\else{\ifnum#1=
87{{\sl Proceedings of the Nineteenth Annual Symposium
on the Theory of Computing,\/} ACM, 1987}\else{\ifnum#1=
88{{\sl Proceedings of the Twentieth Annual Symposium
on the Theory of Computing,\/} ACM, 1988}\else{\ifnum#1=
89{{\sl Proceedings of the Twenty First Annual Symposium
on the Theory of Computing,\/} ACM, 1989}\else{\ifnum#1=
90{{\sl Proceedings of the Twenty Second Annual Symposium
on the Theory of Computing,\/} ACM, 1990}\else{\ifnum#1=
91{{\sl Proceedings of the Twenty Third Annual Symposium
on the Theory of Computing,\/} ACM, 1991}\else{\ifnum#1=
92{{\sl Proceedings of the Twenty Fourth Annual Symposium
on the Theory of Computing,\/} ACM, 1992}\else{\ifnum#1=
93{{\sl Proceedings of the Twenty Fifth Annual Symposium
on the Theory of Computing,\/} ACM, 1993}\else{\ifnum#1=
94{{\sl Proceedings of the Twenty Sixth Annual Symposium
on the Theory of Computing,\/} ACM, 1994}\else{\ifnum#1=
95{{\sl Proceedings of the Twenty Seventh Annual Symposium
on the Theory of Computing,\/} ACM, 1995}\else{\ifnum#1=
96{{\sl Proceedings of the Twenty Eighth  Annual Symposium
on the Theory of Computing,\/} ACM, 1996}\else{\ifnum#1=
97{{\sl Proceedings of the Twenty Eighth  Annual Symposium
on the Theory of Computing,\/} ACM, 1997}\else{
This STOC not yet defined!}
\fi}\fi}\fi}\fi}\fi}\fi}\fi}\fi}\fi}\fi}\fi}\fi}\fi}\fi}\fi}\fi}\fi}

\newcommand{\focs}[1]{\ifnum#1=
78{{\sl Proceedings of the Nineteenth Annual Symposium
on the Foundations of Computer Science,\/} IEEE, 1978}\else{\ifnum#1=
80{{\sl Proceedings of the Twenty First Annual Symposium
on the Foundations of Computer Science,\/} IEEE, 1980}\else{\ifnum#1=
82{{\sl Proceedings of the Twenty Third Annual Symposium
on the Foundations of Computer Science,\/} IEEE, 1982}\else{\ifnum#1=
85{{\sl Proceedings of the Twenty Sixth Annual Symposium
on the Foundations of Computer Science,\/} IEEE, 1985}\else{\ifnum#1=
86{{\sl Proceedings of the Twenty Seventh Annual Symposium
on the Foundations of Computer Science,\/} IEEE, 1986}\else{\ifnum#1=
87{{\sl Proceedings of the Twenty Eighth Annual Symposium
on the Foundations of Computer Science,\/} IEEE, 1987}\else{\ifnum#1=
88{{\sl Proceedings of the Twenty Ninth Annual Symposium
on the Foundations of Computer Science,\/} IEEE, 1988}\else{\ifnum#1=
89{{\sl Proceedings of the Thirtieth Annual Symposium
on the Foundations of Computer Science,\/} IEEE, 1989}\else{\ifnum#1=
90{{\sl Proceedings of the Thirty First Annual Symposium
on the Foundations of Computer Science,\/} IEEE, 1990}\else{\ifnum#1=
91{{\sl Proceedings of the Thirty Second Annual Symposium
on the Foundations of Computer Science,\/} IEEE, 1991}\else{\ifnum#1=
92{{\sl Proceedings of the Thirty Third Annual Symposium
on the Foundations of Computer Science,\/} IEEE, 1992}\else{\ifnum#1=
93{{\sl Proceedings of the Thirty Fourth Annual Symposium
on the Foundations of Computer Science,\/} IEEE, 1993}\else{\ifnum#1=
94{{\sl Proceedings of the Thirty Fifth Annual Symposium
on the Foundations of Computer Science,\/} IEEE, 1994}\else{\ifnum#1=
95{{\sl Proceedings of the Thirty Sixth  Annual Symposium
on the Foundations of Computer Science,\/} IEEE, 1995}\else{\ifnum#1=
96{{\sl Proceedings of the Thirty Seventh Annual Symposium
on the Foundations of Computer Science,\/} IEEE, 1996}\else{\ifnum#1=
97{{\sl Proceedings of the Thirty Eigth Annual Symposium
on the Foundations of Computer Science,\/} IEEE, 1997}
\else{
This FOCS not yet defined!}
\fi}\fi}\fi}\fi}\fi}\fi}\fi}\fi}\fi}\fi}\fi}\fi}\fi}\fi}\fi}\fi}

\newcommand{\icalp}[1]{\ifnum#1=
78{{\sl Proceedings of ICALP~78,\/} Lecture Notes
in Computer Science Vol.~62, Springer-Verlag, 1978}\else{\ifnum#1=
81{{\sl Proceedings of ICALP~81,\/} Lecture Notes
in Computer Science Vol.~115, Springer-Verlag, 1981}\else{\ifnum#1=
89{{\sl Proceedings of ICALP~89,\/} Lecture Notes in
Computer Science Vol.~372, Springer Verlag, 1989}\else{\ifnum#1=
90{{\sl Proceedings of ICALP~90,\/} Lecture Notes in
Computer Science Vol.~443, Springer Verlag, 1990}\else{\ifnum#1=
91{{\sl Proceedings of ICALP~91,\/} Lecture Notes in
Computer Science Vol.~510, Springer Verlag, 1991}\else{\ifnum#1=
92{{\sl Proceedings of ICALP~92,\/} Lecture Notes in
Computer Science Vol.~623, Springer Verlag, 1992}\else{\ifnum#1=
93{{\sl Proceedings of ICALP~93,\/} Lecture Notes in
Computer Science Vol.~700, Springer Verlag, 1993}\else
This ICALP not yet defined!
\fi}\fi}\fi}\fi}\fi}\fi}\fi}

\newcommand{\istcs}[1]{\ifnum#1=
93{{\sl Proceedings of the Second Israel Symposium on Theory
and Computing Systems\/}, 1993}\else{\ifnum#1=
95{{\sl Proceedings of the Third Israel Symposium on Theory
and Computing Systems\/}, 1995}\else
This ISTCS not yet defined!
\fi}\fi}

\newcommand{\fsttcs}[1]{\ifnum#1=
94{{\sl Proceedings of the Fourteenth Annual Symposium on
Foundations of Software Technology and
Theoretical Computer Science,\/} Lecture Notes in Computer
Science Vol.~880, Springer Verlag, 1994} \else
This FSTTCS not yet defined!
\fi}

\newcommand{\stacs}[1]{\ifnum#1=
90{{\sl Proceedings of the Seventh Annual Symposium on
Theoretical Aspects of Computer Science,\/} Lecture Notes in
Computer Science Vol.~415, Springer Verlag, 1990}\else{\ifnum#1=
91{{\sl Proceedings of the Eighth Annual Symposium on
Theoretical Aspects of Computer Science,\/} Lecture Notes in
Computer Science Vol.~480, Springer Verlag, 1991}\else{\ifnum#1=
92{{\sl Proceedings of the Ninth Annual Symposium on
Theoretical Aspects of Computer Science,\/} Lecture Notes in
Computer Science Vol.~577, Springer Verlag, 1992}\else{\ifnum#1=
93{{\sl Proceedings of the Tenth Annual Symposium on
Theoretical Aspects of Computer Science,\/} Lecture Notes in
Computer Science Vol.~665, Springer Verlag, 1993}\else
This STACS not yet defined!
\fi}\fi}\fi}\fi}

\newcommand{\structures}[1]{\ifnum#1=
88{{\sl Proceedings of the Third Annual Conference on
Structure in Complexity Theory\/}, IEEE, 1988}\else{\ifnum#1=
89{{\sl Proceedings of the Fourth Annual Conference on
Structure in Complexity Theory\/}, IEEE, 1989}\else{\ifnum#1=
90{{\sl Proceedings of the Fifth Annual Conference on
Structure in Complexity Theory\/}, IEEE, 1990}\else{\ifnum#1=
91{{\sl Proceedings of the Sixth Annual Conference on
Structure in Complexity Theory\/}, IEEE, 1991}\else{\ifnum#1=
92{{\sl Proceedings of the Seventh Annual Conference on
Structure in Complexity Theory\/}, IEEE, 1992}\else{\ifnum#1=
93{{\sl Proceedings of the Eighth Annual Conference on
Structure in Complexity Theory\/}, IEEE, 1993}\else{\ifnum#1=
94{{\sl Proceedings of the Ninth Annual Conference on
Structure in Complexity Theory\/}, IEEE, 1994}\else{\ifnum#1=
96{{\sl Proceedings of the Eleventh Annual Conference on
Complexity Theory\/}, IEEE, 1996}\else{\ifnum#1=
97{{\sl Proceedings of the Twelfth Annual Conference on
Complexity Theory\/}, IEEE, 1997}\else
This Structures not yet defined!
\fi}\fi}\fi}\fi}\fi}\fi}\fi}\fi}\fi}

\newcommand{\soda}[1]{\ifnum#1=
92{{\sl Proceedings of the Third Symposium on Discrete
Algorithms\/}, ACM, 1994}\else{\ifnum#1=
93{{\sl Proceedings of the Third Symposium on Discrete
Algorithms\/}, ACM, 1993}\else{\ifnum#1=
94{{\sl Proceedings of the Fifth Symposium on Discrete
Algorithms\/}, ACM, 1994}\else{\ifnum#1=
95{{\sl Proceedings of the Sixth Symposium on Discrete
 Algorithms\/}, ACM, 1995}\else
This SODA not yet defined!
\fi}\fi}\fi}\fi}

\label{lastpage}


\begin{thebibliography}{188}
%\bibitem%
%{arora:thesis}
%{ S.~Arora.}
%{\em Probabilistic Checking of Proofs and Hardness of Approximation
%Problems.}
%PhD thesis, U.C. Berkeley, 1994.
%Available from {\tt http://www.cs.princeton.edu/\~{ }arora }.

\bibitem{arlu}
{ S.~Arora and C.~Lund.}
\newblock Hardness of approximations.
\newblock In~\cite{hochbook}.

\bibitem%[ALM{\etalchar{+}}92]
{ALMSS}
S.~Arora, C.~Lund, R.~Motwani, M.~Sudan, and M.~Szegedy.
\newblock Proof verification and intractability of approximation problems.
\newblock In {\em Proc. 33rd IEEE Symp. on Foundations of Computer Science},
  13--22, 1992.

%\bibitem%[AMSSS92]
%{amsss}
%{ S.~Arora, R.~Motwani, S.~Safra, M.~Sudan, and M.~Szegedy.}
%\newblock PCP and approximation problems.
%\newblock {\em Unpublished note}, 1992.

\bibitem%[ArSa]
{arsa}
{ S. Arora and S. Safra}.
Probabilistic checking of proofs:~a new characterization of NP.
To appear {\em Journal of the ACM}.
Preliminary version in \focs{92}.

\bibitem{arsu}
{ S. Arora and M. Sudan}.
Improved low degree testing and its applications.
\stoc{97}


\bibitem{babai}%[Bab85]
L.~Babai.
\newblock Trading group theory for randomness.
\newblock In {\em Proc. 17th ACM Symp. on Theory of Computing}, 421--429,
  1985.

\bibitem{babaisurvey}%[Bab94]
L.~Babai.
\newblock Transparent proofs and limits to approximations.
\newblock In {\em Proceedings of the First European Congress of
  Mathematicians}. Birkhauser, 1994.

\bibitem{BFL}%[BFL91]
L.~Babai, L.~Fortnow, and C.~Lund.
\newblock Non-deterministic exponential time has two-prover interactive
  protocols.
\newblock {\em Computational Complexity}, 1:3--40, 1991.

\bibitem{BFLS}%[BFLS91]
L.~Babai, L.~Fortnow, L.~Levin, and M.~Szegedy.
\newblock Checking computations in polylogarithmic time.
\newblock In {\em Proc. 23rd ACM Symp. on Theory of Computing}, 21--31,
  1991.

\bibitem{bamo}%[BaM]
L.~Babai and S.~Moran.
Arthur-Merlin games: a randomized proof system,
and a hierarchy of complexity classes.
{\em Journal of Computer and System Sciences},
36:254-276, 1988.

\bibitem{barak}
B. Barak.
\newblock How to Go Beyond the Black-Box Simulation Barrier.
In {\em Proc. 42nd IEEE FOCS}, 106--115, 2001.

\bibitem{befe}%[BeF]
{ D.~Beaver and J.~Feigenbaum.}
Hiding instances in multioracle queries.
\stacs{90}.


\bibitem{bchks}%[BCHKS]
{ M.~Bellare, D.~Coppersmith, J.~H{\aa}stad, M.~Kiwi and M.~Sudan.}
Linearity testing in characteristic two.
{\em IEEE Transactions on Information Theory\/}
42(6):1781-1795, November 1996.

\bibitem{BGS}%[BGS95]
M.~Bellare, O.~Goldreich, and M.~Sudan.
\newblock Free bits and non-approximability-- towards tight results.
\newblock In {\em Proc. 36th IEEE Symp. on Foundations of Computer Science},
  1995.
\newblock Full version available from ECCC.

\bibitem{BGLR}%[BGLR93]
M.~Bellare, S.~Goldwasser, C.~Lund, and A.~Russell.
\newblock Efficient multi-prover interactive proofs with applications to
  approximation problems.
\newblock In {\em Proc. 25th ACM Symp. on Theory of Computing},  113--131,
  1993.


\bibitem{BS}%[BS94]
M.~Bellare and M.~Sudan.
\newblock Improved non-approximability results.
\newblock In {\em Proc. 26th ACM Symp. on Theory of Computing},  184--193,
  1994.

\bibitem{BGKW}%[BGKW88]
M.~{Ben-or}, S.~Goldwasser, J.~Kilian, and A.~Wigderson.
\newblock Multi prover interactive proofs: How to remove intractability
  assumptions.
\newblock In {\em Proc. 20th ACM Symp. on Theory of Computing}, 113--121,
  1988.


\bibitem{BK}%[BK89]
M.~Blum and S.~Kannan.
\newblock Designing programs that check their work.
\newblock {\em JACM} {\bf 42}(1):269-291, 1995.
%In {\em Proc. 21st ACM Symp. on Theory of Computing}, 86--97,
%  1989.

\bibitem{BLR}
M.~Blum, M.~Luby, and R.~Rubinfeld.
\newblock Self-Testing/Correcting with Applications to Numerical
                 Problems.
\newblock {\em JCSS} 47(3):549--595, 1993.
%\newblock {\em Prelim. Version in} \stoc{90}.

\bibitem{clayprobs}
Clay Institute.
\newblock Millenium Prize Problems.
%\newblock http://www.claymath.org/prizeproblems/index.htm

\bibitem{condon}%[Con93]
A.~Condon.
\newblock The complexity of the max-word problem and the power of one-way
  interactive proof systems.
\newblock {\em Computational Complexity}, 3:292--305, 1993.

\bibitem%[CFLS]
{cfls}
{ A.~Condon, J.~Feigenbaum, C.~Lund and P.~Shor.}
Probabilistically Checkable Debate Systems and Approximation
Algorithms for PSPACE-Hard Functions.
\stoc{93}.

\bibitem%[CFLS]
{cfls2}
{ A.~Condon, J.~Feigenbaum, C.~Lund and P.~Shor.}
Random debaters and the hardness of approximating stochastic
functions.
{\em SIAM J. Comp.},
26(2):369-400, 1997.

\bibitem{cook}%[Coo71]
S.~Cook.
\newblock The complexity of theo\-rem-proving procedures.
\newblock In {\em Proc. 3rd ACM Symp. on Theory of Computing}, 151--158,
  1971.


\bibitem{CreKann}%[CK94]
P.~Crescenzi and V.~Kann.
\newblock A compendium of {NP} optimization problems.
\newblock Available from ftp://www.nada.kth.se/Theory/Viggo-Kann/compendium.ps.Z

\bibitem%[Fei95]
{uri:setcover}
U.~Feige.
\newblock A threshold of $\ln n$ for approximating set cover.
\newblock \stoc{96}, 314--318.


\bibitem{FGLSS}%[FGL{\etalchar{+}}91]
U.~Feige, S.~Goldwasser, L.~Lov\'asz, S.~Safra, and M.~Szegedy.
\newblock Interactive proofs and the hardness of approximating
                cliques
\newblock {\em Journal of the ACM}, 43(2):268--292, 1996.
%\newblock {\em Preliminary version:
%Approximating clique is almost {NP}-complete, \focs{91}.}

\bibitem{FKcli}%[FK94]
U.~Feige and J.~Kilian.
\newblock Two prover protocols--low error at affordable rates.
\newblock In {\em Proc. 26th ACM Symp. on Theory of Computing},  172--183,
  1994.


\bibitem{FL}%[FL92]
U.~Feige and L.~Lov\'asz.
\newblock Two-prover one-round proof systems: Their power and their problems.
\newblock In {\em Proc. 24th ACM Symp. on Theory of Computing},  733--741,
  1992.

\bibitem{FRS}
%[FRS88]
L.~Fortnow, J.~Rompel, and M.~Sipser.
\newblock On the power of multi-prover interactive protocols.
\newblock In {\em Proceedings of the 3rd Conference on Structure in Complexity
  Theory},  156--161, 1988.


\bibitem%[GJ79]
{GarJon}
M.~R. Garey and D.~S. Johnson.
\newblock {\em Computers and Intractability: a guide to the theory of
  {NP}-completeness}.
\newblock W. H. Freeman, 1979.

\bibitem%[Gol94]
{G}
O.~Goldreich.
\newblock Probabilistic proof systems.
%\newblock Technical Report RS-94-28, Basic Research in Computer Science, Center
%  of the Danish National Research Foundation, September 1994.
\newblock {\em Proceedings of the International Congress of
  Mathematicians}, Birkhauser Verlag 1994.


\bibitem{goldreich}
{ O. Goldreich.}
A Taxonomy of Proof Systems.
In {\em Complexity Theory Retrospective II},
L.A.~Hemaspaandra and A.~Selman (eds.),
Springer-Verlag, New York, 1997.


\bibitem{GGR}
O.~Goldreich, S.~Goldwasser and D.~Ron.
\newblock Property Testing and its Connection to Learning and
                Approximation.
\newblock \focs{96},  339--348.

\bibitem%[GMR89]
{GMR}
S.~Goldwasser, S.~Micali, and C.~Rackoff.
\newblock The knowledge complexity of interactive proofs.
\newblock {\em SIAM J. on Computing}, 18:186--208, 1989.

%\bibitem%[GMW87]
%{GMW}
%O.~Goldreich, S.~Micali, and A.~Wigderson.
%\newblock How to play any mental game or a completeness theorem for protocols
%  with honest majority.
%\newblock In {\em Proc. 19th ACM Symp. on Theory of Computing},  218--229,
%  1987.

\bibitem{GS}
V.~Guruswami and M.~Sudan.
\newblock Improved decoding of Reed-Solomon and Algebraic-Geometric codes.
\newblock {\em IEEE Trans.\  Inf.\ Thy.}, {\bf 45}:1757-1767 (1999).


\bibitem%[Has95]
{hastadcli}
J.~H{\aa}stad.
\newblock Clique is Hard to Approximate within $n^{1-\epsilon}$.
\newblock {\em Acta Mathematica}, {\bf 182}:105-142, 1999.

\bibitem{hastadmtsat}
J.~H{\aa}stad.
\newblock Some optimal inapproximability results.
\newblock \stoc{97},  1--10.


\bibitem{hochbook}
D.~Hochbaum, ed.
\newblock Approximation Algorithms for NP-hard problems.
\newblock PWS Publishing, Boston, 1996.


\bibitem%[Joh92]
{johnsonsurvey}
D.~S. Johnson.
\newblock The {NP}-completeness column: an ongoing guide.
\newblock {\em Journal of Algorithms}, 13:502--524, 1992.

\bibitem{KZ}
H.~Karloff and U.~Zwick.
\newblock A 7/8-Approximation Algorithm for {MAX} 3{SAT}?
\newblock In {\em Proc. IEEE FOCS}, 1997.

\bibitem%[Kar72]
{karp72}
R.~M. Karp.
\newblock Reducibility among combinatorial problems.
\newblock In Miller and Thatcher, editors, {\em Complexity of Computer
  Computations}, 85--103. Plenum Press, 1972.

\bibitem{kilian}
J.~Kilian.
\newblock Improved Efficient Arguments.
\newblock {\em Advances in Cryptology - CRYPTO 95}
LNCS 963 Coppersmith, D.(ed.), Springer, NY, 311-324, 1995.

\bibitem{klrss}
M.~Kiwi, C.~Lund, A.~Russell, D.~Spielman, and R.~Sundaram
\newblock Alternation in interaction.
\structures{94}, 294--303.

\bibitem{kolata}
G.~Kolata.
\newblock New shortcut  found for long math proofs.
\newblock {\em New York Times}, April 7, 1992.


\bibitem%[LS91]
{LS}
D.~Lapidot and A.~Shamir.
\newblock Fully parallelized multi prover protocols for {NEXPTIME}.
\newblock In {\em Proc. 32nd IEEE Symp. on Foundations of Computer Science},
   13--18, 1991.


\bibitem%[Lev73]
{levin}
L.~Levin.
\newblock Universal'ny\u{\i}e pereborny\u{\i}e zadachi (universal search
  problems : in {R}ussian).
\newblock {\em Problemy Peredachi Informatsii}, 9(3):265--266, 1973.


\bibitem{lipton}
R.~Lipton.
New directions in testing.
\newblock In {\em Distributed Computing and Cryptography},
J. Feigenbaum and M. Merritt, eds.
\newblock Dimacs Series in Discrete Mathematics and Theoretical
Computer Science, 2. AMS 1991.

\bibitem%[LFKN92]
{LFKN}
C.~Lund, L.~Fortnow, H.~Karloff, and N.~Nisan.
\newblock Algebraic methods for interactive proof systems.
\newblock {\em JACM}, 39(4):859--868, 1992.

\bibitem%[LY94]
{LY}
C.~Lund and M.~Yannakakis.
\newblock On the hardness of approximating minimization problems.
\newblock {\em JACM}, 41(5):960--981, 1994.

\bibitem{MPS}
E.~W.~Mayr, H.~J.~Promel, and A.~Steger. (eds)
\newblock Lectures on proof verification and approximation algorithms.
\newblock LNCS Tutorial, Springer Verlag, 1998.

\bibitem{micaliCS}
S.~Micali.
\newblock Computationally Sound Proofs.
\newblock {\em SIAM J.~Comp}, {\bf 30}(4), 1253--1298.

\bibitem%[PY91]
{PY}
C.~Papadimitriou and M.~Yannakakis.
\newblock Optimization, approximation and complexity classes.
\newblock {\em Journal of Computer and System Sciences}, 43:425--440, 1991.
%\newblock {\em Prelim.~version in Proc. ACM STOC 1988.}


\bibitem%[Raz94]
{raz}
R.~Raz.
\newblock A parallel repetition theorem.
\newblock {\em SIAM J.~Comp} 27(3):763-803(1998).


\bibitem%[RS97]
{razsaf}
{ R. Raz and S. Safra}.
A sub-constant error-probability low-degree test,
and a sub-constant error-probability PCP characterization of NP.
\stoc{97}.

\bibitem{dron}
D.~Ron.
\newblock Property testing (A Tutorial). In,
{\em Handbook of Randomized Algorithms},
Pardalos, Rajasekaran, Reif, and Rolim, eds., . Kluwer Academic, 2001.


\bibitem%[RS92]
{RS}
R.~Rubinfeld and M.~Sudan.
\newblock Testing polynomial functions efficiently and over rational domains.
\newblock In {\em Proc. 3rd Annual ACM-SIAM Symp. on Discrete Algorithms},
   23--32, 1992.

\bibitem{STZ}
S.~Safra, A.~Ta-Shma, and D.~Zuckerman.
\newblock Extractors from Reed-Muller codes.
\newblock In {\em Proc. IEEE FOCS}, 2001.

\bibitem{ST}
A.~Samorodnitsky and L.~Trevisan.
\newblock A PCP Characterization of NP with Optimal Amortized Query Complexity.
\newblock In {\em Proc. 32nd ACM STOC}, 2000.

\bibitem%[Sha92]
{Sha}
A.~Shamir.
\newblock {IP = PSPACE}.
\newblock {\em JACM}, 39(4):869--877, October 1992.

\bibitem{STV}
M.~Sudan, L.~Trevisan and S.~Vadhan.
\newblock Pseudorandom Generators Without the XOR Lemma.
\newblock {\em JCSS}, {\bf 62}(2):236-266, 2001.

\bibitem{trevisan}
Luca Trevisan.
\newblock Extractors and Pseudorandom Generators.
\newblock {\em JACM}, {\bf 48}(4):860-879, 2001.

\bibitem{vijaybook}
Vijay Vazirani.
\newblock Approximation Algorithms.
\newblock Spring Verlag, 2001.

\end{thebibliography}
\end{document}